\documentclass[oldversion]{aa}  
\usepackage{graphicx,psfig,amsmath}
\begin{document}
   \title{Rayleigh scattering in the transit spectrum of HD\,189733b}

   \author{
   A.~Lecavelier des Etangs\inst{1,2}
   \and
   F.~Pont\inst{3}          
   \and
   A.~Vidal-Madjar\inst{1,2}
   \and 
   D.~Sing\inst{1,2}
}
   
   \authorrunning{}%A. Lecavelier et al.}

   \offprints{A.L. (\email{lecaveli@iap.fr})}

   \institute{
   CNRS, UMR 7095, 
   Institut d'Astrophysique de Paris, 
   98$^{\rm bis}$ boulevard Arago, F-75014 Paris, France
   \and
   UPMC Univ. Paris 6, UMR 7095, 
   Institut d'Astrophysique de Paris, 
   98$^{\rm bis}$ boulevard Arago, F-75014 Paris, France
   \and
   Physikalisches Institut, University of Bern, Sidlerstrasse 5, 3012 Bern, Switzerland    }   
   \date{Received ...; accepted ...}
 
  \abstract
{
   The transit spectrum of the exoplanet HD\,189733b has recently been obtained between 0.55 and 1.05$\mu$m. 
   Here we present an analysis of this spectrum.
   We develop first-order equations to interpret absorption spectra.
   In the case of HD\,189733b, we show that the observed slope of the absorption as a function 
   of wavelength is characteristic of extinction proportional to the inverse 
   of the fourth power of the wavelength ($\propto\lambda^{-4}$). Assuming an
   extinction dominated by Rayleigh scattering, we derive an atmospheric
   temperature of 1340\,$\pm$\,150\,K. 
   If molecular hydrogen is responsible for the Rayleigh scattering, the atmospheric
   pressure at the planetary characteristic radius of 0.1564 stellar radius 
   must be 410\,$\pm$\,30\,mbar. However the preferred scenario is 
   scattering by condensate particles. Using the Mie approximation,
   we find that the particles must have a low value 
   for the imaginary part of the refraction index. We identify MgSiO$_3$ as a 
   possible abundant condensate whose particle size must be between $\sim$$10^{-2}$ 
   and $\sim$$10^{-1}$\,$\mu$m.
   For this condensate, assuming solar abundance, the pressure at 0.1564~stellar radius 
   is found to be between a few microbars and few millibars, and the temperature is found
   to be in the range 1340-1540\,K, and both depend on the particle size.   
}

   \keywords{Stars: planetary systems}

%\authorrunning{Lecavelier et al.}
\titlerunning{Rayleigh scattering in HD\,189733b}

   \maketitle
%
%________________________________________________________________

\section{Introduction}
\label{Introduction}

Transiting extrasolar planets, like HD\,189733b offer unique opportunities
to scrutinize their atmospheric content 
(e.g., Charbonneau et al.\ 2002, Vidal-Madjar et al.\ 2003). 
In particular, primary transits can reveal tenuous quantities of gas or dust 
by spectral absorption leading to variations in the apparent planet size
as a function of the wavelength. 
HD\,189733\,b is presently the best target for such studies because of
its nearby and bright host star, its large absorption depth 
in its transit light curve (Bouchy et al.\ 2005),
and a short period (H\'ebrard \& Lecavelier des Etangs 2006).

Using transit observations
with the ACS camera of the Hubble Space Telescope,
the apparent radius of 
HD\,189733b has been measured 
from 0.55 to 1.05 microns, 
thus providing the ``transmission spectrum'' of the atmosphere
(Pont et al.\ 2008).
In this wavelength range, strong features of abundant atomic 
species and molecules were 
predicted but not detected, leading Pont et al.\ (2008)
to the conclusion that a haze of sub-micron particles 
is present in the upper atmosphere of the planet.

Here we use these measurements of the planet radius
as a function of the wavelength from 550 to 1050\,nm. 
In Sect.~\ref{Interpreting} we describe how a transmission spectrum
can be interpreted to derive basic quantities.
We then estimate the temperature of HD\,189733b from
fit to the ACS spectrum and show that the observed absorption
can be caused by Rayleigh scattering (Sect.~\ref{Temperature}). 
Several species that are possibly responsible for the Rayleigh scattering 
are discussed, and the corresponding pressure at the 
absorption level are estimated in Sect.~\ref{Rayleigh}.

\section{Interpreting a transit spectrum}
\label{Interpreting}

\subsection{Planet radius as a function of atmospheric content}

\begin{figure}[bth!]
%\hspace{0.5cm}
\psfig{file=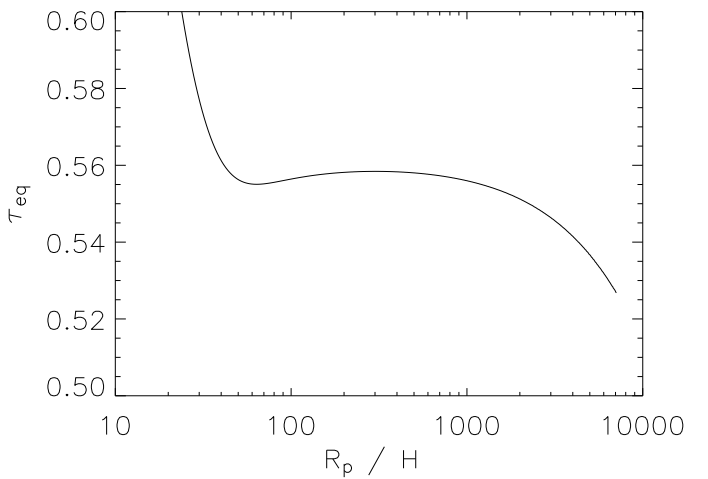,width=\columnwidth,angle=90} 
\vspace{0.4cm}
\caption[]{
Plot of $\tau_{eq}$ as a function of the ratio of the planet radius to the atmosphere scale height.
\label{Tau_eq}}
\end{figure}

To interpret transit spectra, one can use detailed atmosphere 
models that include temperature, pressure, and composition as a function
of the altitude and numerically integrate the radiative transfer 
equations to obtain a theoretical spectrum to compare to the observations
(e.g., Ehrenreich et al.\ 2006). Here we propose another approach
by deriving first-order equations to obtain a better feeling
for the basic quantities that can be obtained from these
measurements.

Following Fortney's (2005) derivation of the path length, 
the optical depth, $\tau$, 
in a line of sight grazing the planetary limb at an altitude $z$ 
is given by
$\tau(\lambda,z)\approx \sigma(\lambda)n(z)\sqrt{2\pi R_{planet} H}$, 
where $R_{planet}$ is the assumed planet radius, 
$H$ the atmosphere scale height, and $n(z)=n_{(z=0)} exp(-z/H)$ the
volume density at the altitude $z$ of the main absorbent 
with a cross section $\sigma(\lambda)$. 
For a temperature $T$, $H$ is given by $ H = k T/\mu g$,
where $\mu$ is the mean mass of atmospheric particles taken to 
be 2.3 times the mass of the proton, and $g$ the gravity.
Using planetary transit measurements, a fit to the light curve
provides the ratio of the effective 
planetary radius as a function of wavelength to
the stellar radius: $R_p(\lambda)/R_*$. 
Here we define $\tau_{eq}$ by the optical depth at altitude $z_{eq}$ 
such that a sharp occulting disk of radius $R_{planet}+z_{eq}$ produces 
the same absorption depth as the planet with its translucent atmosphere; 
in other words, $\tau_{eq}$ is defined by 
$R_p(\lambda) = R_{planet}+z(\tau= \tau_{eq})$. 
Using a model atmosphere, and numerically integrating over 
the whole translucent atmosphere, we can calculate the 
effective planet radius, and then obtain the corresponding 
optical thickness at the effective radius, $\tau_{eq}$. 
For various atmospheric scale heights, 
we calculate $\tau_{eq}$ by numerical integration. 
For a wide range of atmospheric
scale height, provided that R$_p$/H is between $\sim$30 and $\sim$3000,
the resulting $\tau_{eq}$ is roughly constant at a value 
$\tau_{eq}\approx 0.56$ (Fig.~\ref{Tau_eq}).
In the case of HD\,189733b, R$_p$/H varies between 280 and 560 
when the temperature varies from 1000 to 2000 K; therefore,
the approximation of a constant $\tau_{eq}$ at 0.56 fully applies.
This demonstrates that, for a given atmospheric structure 
and composition, estimating the altitude 
at which $\tau=\tau_{eq}=0.56$ is all that is needed to calculate
the effective radius of the planet at a given wavelength.

\subsection{Consequences}
\label{Consequences}

For a given atmospheric structure and composition, 
the effective altitude of the atmosphere
at a wavelength $\lambda$ is calculated by finding $z(\tau = \tau_{eq} ,\lambda )$,
which solves the equation $\tau(\lambda,z)=\tau_{eq}$.
Using the quantities defined above, the effective altitude $z$ is given by 
\begin{equation}
z(\lambda)=H\ln \left(\xi_{abs} P_{z=0} \sigma_{abs}(\lambda) / \tau_{eq}
 \times \sqrt{2\pi  R_p/kT \mu g}\right),
\label{z_lambda}
\end{equation}
where $\sigma_{abs}$ and $\xi_{abs}$ 
are the cross section and abundance of the dominant absorbing species.

If the variation of the cross section as a function of wavelength is known, the
observation of the altitude as a function of wavelength allows the derivation of
$H$ and, therefore, of the temperature $T$, given by:
\begin{equation}
T=\frac{\mu g}{k} \left( \frac{d\ln \sigma}{d\lambda}\right)^{-1}
\frac{d z(\lambda)}{d\lambda}.
\label{T_dz}
\end{equation}

Using Eq.~\ref{z_lambda}, 
the partial pressure of the main absorbent at the reference altitude
is estimated by 
\begin{equation}
\xi_{abs} P_{z=0} = \tau_{eq}/ \sigma_{abs}(\lambda_{z=0}) \times \sqrt{kT \mu g /2\pi  R_p}, 
\label{xiP}
\end{equation}
where $\lambda_{z=0}$ is the wavelength at which the effective 
planetary radius corresponds to (or is used to define) $z=0$.
We note that there is a degeneracy between the abundance and the total pressure
in the atmosphere. From the measurement of the effective radius using
transit spectroscopy, one needs to assume an abundance of the absorbent 
to derive the pressure, or alternatively to assume a pressure to derive the abundance.

In summary, the effective planetary radius is characteristic
of the pressure and abundance, and the variation in this
radius as a function of wavelength is characteristic
of the temperature. 

\section{The atmospheric temperature in HD\,189733b}
\label{Temperature}

\begin{figure}[tb!]
%\hspace{0.5cm}
\psfig{file=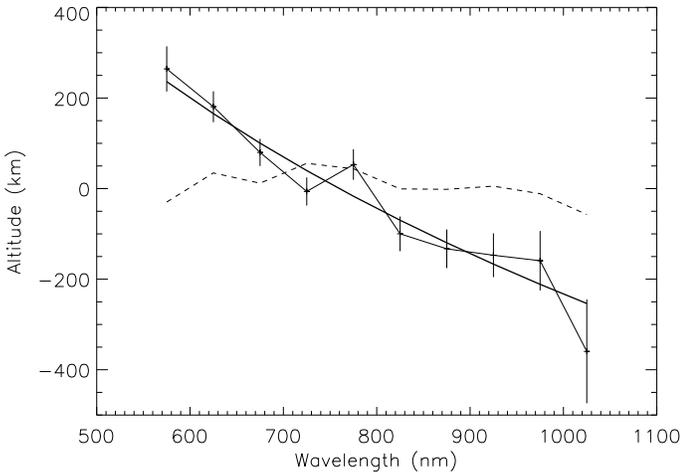,width=\columnwidth,angle=90} 
%\vspace{0.5cm}
%\resizebox{\hsize}{!}{\includegraphics{H_vs_Lambda.ps}}
\caption[]{
Plot of the altitude in the atmosphere of HD189733\,b corresponding to the measured planetary
radius as a function of wavelength. Here the zero altitude is defined for a planetary radius
of 0.1564 times the stellar radius. The measured values with error bars 
are those of Pont et al.\ (2008). The fit to the data (thick line) is obtained assuming an
extinction proportional to $\lambda^{-4}$ and for
an atmospheric temperature of 1340\,K.
Fits using H$_2$ Rayleigh scattering or MgSiO$_3$ grains with 
sizes between 0.01 and 0.1\,$\mu$m give the same plots.
Larger grain diameters are not consistent with the observations
({\it e.g.}, 0.4\,$\mu$m, shown with dotted line).
\label{H_vs_Lambda}}
\end{figure}

In HD\,189733b, the plot of altitude as a function 
of the wavelength shows 
an increase in absorption toward shorter wavelengths 
(Fig.~\ref{H_vs_Lambda}). 
Using Eq.~\ref{T_dz} 
and assuming a scaling law for the cross section in the form
$\sigma= \sigma_0 ( \lambda/ \lambda_0)^{\alpha}$,
the slope of the planet radius as a function of the wavelength 
is given by 
$
dR_p/d\ln\lambda = dz/d\ln\lambda = \alpha H.
$
Therefore, we have 
\begin{equation}
\alpha T= \frac{\mu g}{k} \frac{dR_p}{d\ln\lambda}.
\label{T_alpha}
\end{equation}

For HD\,189733b, the slope is measured to be 
472\,km\,$\pm$42\,(stat.)\,$\pm$106\,(syst.) 
from 600 to 1000\,nm (Pont et al.\ 2008).
This corresponds to $dR_p/d\ln\lambda$\,$\approx$\,$-920$\,km. 
We therefore obtain $\alpha T$\,$\approx$\,$-5840$\,K.
The temperature in HD\,189733b is determined to be in the range 900--1500\,K
(Deming et al.\ 2006; Knutson et al.\ 2007);
therefore, $\alpha$ is found to be close to 
$\alpha\approx -4$,
which is typical of Rayleigh scattering.

Assuming $\alpha=-4$ as expected for Rayleigh scattering 
and using Eq.~\ref{T_dz}, we derive a temperature 
$$
T=1460\pm 130 ({\rm stat.}) \pm 330 ({\rm syst.}) \, {\rm K},
$$
where (stat.) and (syst.) correspond to statistical 
and systematics error bars on $dR_p$, as quoted in Pont et al.\ (2008).

For condensates, the scale height $H_c$ is often found to be
significantly smaller than the gaseous scale height $(H_c\sim H /3$)
(Ackerman \& Marley 2001; Fortney 2005).
If this were the case, because of the observed slope in the spectrum, 
the temperature (or $\alpha$) should be about three times higher
than found above, which seems unlikely. 
We therefore conclude that, if produced by dust condensates, the observed transit
spectrum of HD189733\,b shows that these condensates must be
unusually well-mixed vertically with the atmospheric gas
with a similar scale height. 

\section{Rayleigh scattering}
\label{Rayleigh}

\subsection{Molecular hydrogen}

One possible carrier of the Rayleigh scattering is the most abundant molecule, 
molecular hydrogen.
The effective altitude at which the Rayleigh scattering of molecular hydrogen dominates 
only depends on the mean density. Therefore, measurements of the altitude
in the regime where Rayleigh scattering by H$_2$ dominates over other absorbents 
allows the determination of not only the temperature, but because the abundance 
of H$_2$ is close to 1 ($\xi_{H_2}\sim 1$) also the 
total density and consequently the total pressure at the reference zero 
altitude. 

Following Eq.~\ref{xiP}, the pressure $P_0$ at the altitude corresponding 
to the radius at wavelength $\lambda_0$ is 
$$
P_0=\tau_{eq}/ \sigma_0 \times \sqrt{kT \mu g /2\pi  R_p},
$$
where $\sigma_0$ is the Rayleigh scattering cross section at $\lambda_0$.
Using the refractive index of molecular hydrogen $(r-1)=1.32\times 10^{-4}$, 
one derives $\sigma_0=2.52\times 10^{-28}$\,cm$^{2}$ at $\lambda_0=750$\,nm
(Bates 1984; Naus \& Ubachs 2000).
Using these numbers one finds $P_0\approx 400$\,mbar.

A fit with two free parameters (temperature and pressure) 
and assuming that Rayleigh scattering by molecular hydrogen dominates 
leads to a very good fit to the data published by Pont et al.\ (2008) 
with a $\chi^2$ equals 9.8 for 8 degrees of freedom. 
We thus obtain a temperature of 
$T$=1340\,$\pm$\,150\,K and a pressure at $z=0$ (defined by $R_p/R_*=0.1564$) 
of $P$=410\,$\pm$\,30\,mbar.

\subsection{Condensate}

\begin{figure}[tb!]
%\hspace{0.5cm}
\psfig{file=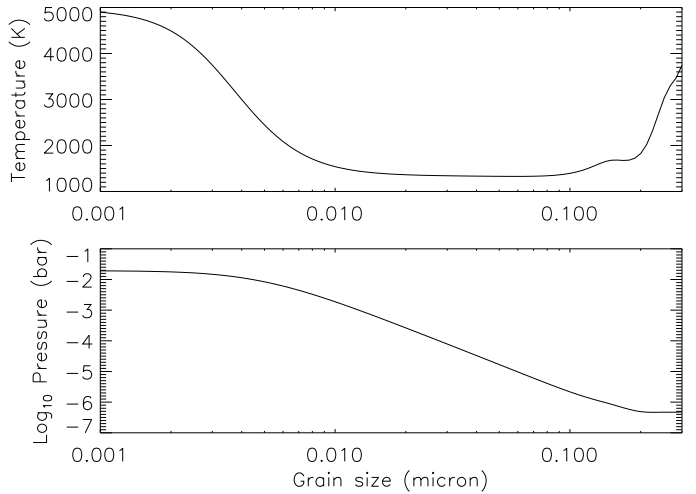,width=\columnwidth,angle=90} 
%\vspace{0.5cm}
%\resizebox{\hsize}{!}{\includegraphics{H_vs_Lambda.ps}}
\caption[]{
Plot of the temperature and pressure at 
zero altitude
as a function of the size of MgSiO$_3$ grains needed to obtain a satisfactory
fit to the data as in Fig.~\ref{H_vs_Lambda}. Here we assume
solar abundance for magnesium atoms to derive the grain abundance.
For particles of 0.03\,$\mu$m, temperature is found to be
the same as with H$_2$ Rayleigh scattering : $T$=1340\,$\pm$\,150\,K.
At 0.01 and 0.1\,$\mu$m temperature increases slightly to 
$T$=1540\,$\pm$\,170\,K and $T$=1390\,$\pm$\,150\,K, respectively.
For particles smaller than 
$\sim$0.01\,$\mu$m and larger than $\sim$0.1\,$\mu$m, Rayleigh scattering
does not dominate the extinction 
($-\alpha$$\ll$$4$), and
the temperature must be higher
to fit the slope of the measured radius as a function of wavelength.
At 0.3\,$\mu$ the fit is bad with $\chi^2$=22.3 for 8 degrees of 
freedom; $\chi^2$ increases steeply for larger sizes.
\label{TP_vs_size}}
\end{figure}

Another possible carrier of the Rayleigh scattering is haze condensate
typically smaller than the wavelength.

Assuming that the absorption measured between 550 and 1050 nm 
is due to particle condensate, we calculate the transmission 
spectrum using the Mie approximation.
In transmission spectroscopy, given the small solid angle
defined by the star seen from the planet, we can use the 
single scattering approximation.  
The glow produced by off-axis scattering is also negligible
(Hubbard et al. 2001).
In this configuration, the Mie extinction efficiency ($Q_{ext}$)
is used and calculated as the summation of the absorption
efficiency ($Q_{abs}$) and scattering efficiency ($Q_{sca}$).

Rayleigh scattering in the form $\lambda^{-4}$ is obtained only
for particle sizes ($a$) much smaller than the wavelength 
($2\pi a$$\ll$$ \lambda $).
In this last case, the absorption and scattering efficiencies 
are given by the approximation:
$Q_{abs}=(8\pi a/\lambda)\Im ((\vec{n}^2-1)/(\vec{n}^2+2))$, 
and 
$Q_{sca}=8/3(2\pi a/\lambda)^4\Re (((\vec{n}^2-1)/(\vec{n}^2+2))^2)$,
where $\vec{n}=n+ik$ is the complex refraction index of the condensate.

We see from the two last equations that Rayleigh scattering 
%(absorption in the form $\lambda^{-4}$) 
dominates the transit spectrum only if $Q_{abs}\ll Q_{sca}$.
Assuming that the imaginary part of the refraction index is much smaller
than the real part, we find that this condition occurs for particle
size, $a$, larger than a minimum size:
$$
a\gg a_{min} = 0.331 \sqrt[3]{\frac{nk}{(n^2-1)^2}}\ \lambda.
$$

For many plausible dust condensates
(like Mg$_2$SiO$_4$, MgFeSiO$_4$, O-deficient silicates, etc.), 
the imaginary part of the refraction 
index is about $k\sim 0.1$, and the real part is in the range
$n\sim 1.5-2.0$ (Dorschner et al.\ 1995). 
Using these values, we find that the minimum particle size is
in the range $a_{min}=\lambda/11$ -- $\lambda/7 $. 
This provides very few possibilities for the particle size,
which must be much smaller than $\lambda/2\pi$ and much larger than
$\sim$$\lambda/10$. We therefore consider unlikely that the transit
spectrum of HD\,189733b can be due to particles with $k\sim 0.1$,
which would require fine-tuning of the particle size 
for the Rayleigh scattering dominating between 550 and 1050\,nm.

A broader range of particle size is allowed if the imaginary
part of the refraction index is significantly smaller than $\sim$0.1.
In other words, the condensate must be transparent enough for the 
absorption to be negligible relative to scattering. 
Among the possible abundant condensates, MgSiO$_3$ presents 
such a property (Dorschner et al.\ 1995; Fortney 2005).
If MgSiO$_3$ contains no Fe atoms, the imaginary part of the
refraction index is found to be $k\sim 2\times 10^{-5}$ in the 
considered wavelength range.  
Using this value, we find $a_{min}/\lambda= 8\times 10^{-3}$.
Therefore, we conclude that MgSiO$_3$ is the candidate species possibly
responsible for the observed Rayleigh scattering in the HD\,189733b
spectrum, and its size must be in the range 
10$^{-2}$--10$^{-1}$~microns (see upper panel of Fig.~\ref{TP_vs_size}).

Assuming an abundance for the MgSiO$_3$ grains and using 
Eq.~\ref{xiP},
we can evaluate the pressure at the effective planetary radius
if the Rayleigh scattering by these grains dominates
the transit spectrum.
For MgSiO$_3$, the refraction index is taken to be $n$$\sim$1.6 
in the range 550-1050\,nm (Dorschner et al.\ 1995). 
This gives a cross section $\sigma\approx 1528 a^6/\lambda^4$.
The grain abundance is obtained using solar abundance and 
assuming that it is limited by the available number of magnesium atoms.
Thus the grains have an abundance 
$\xi_{\rm grain}=2\xi_{\rm Mg} \mu_{\rm MgSiO_3}/(\rho_{\rm grain} 4\pi/3 a^3 N m_p)$,
where $\xi_{\rm Mg}\approx 4\times 10^{-5}$ is the magnesium
abundance, $\mu_{\rm MgSiO_3}$ is about 100.4 times the
mass of the proton ($m_p$), $\rho_{\rm grain} =3.2$\,g\,cm$^{-3}$
is the grain density, 
$a$ the particle size, and $N$ Avogadro's number.
Therefore we have $\xi_{\rm grain}=1.0\times 10^{-15}(a/1\,\mu{\rm m})^{-3}$.
Finally, at $z=0$ defined by the effective radius at $\lambda\approx 750$\,nm,
with T=1340\,K
we obtain a pressure $P=2\times 10^{-3}$\,bar if the size of MgSiO$_3$
grains is about 0.01$\mu$m, and
$P=2\times 10^{-6}$\,bar if the size is about 0.1$\mu$m.
These values obtained with Eq.~\ref{xiP}
are similar to those obtained 
with a general fit to the data (Fig.~\ref{TP_vs_size}).
It is noteworthy that these values are consistent with the
condensation pressure of MgSiO$_3$ which is $\sim$10$^{-4}$\,bar at
1400\,K (Lodders 1999).

Finally, Rayleigh scattering by MgSiO$_3$ is preferred to scattering by H$_2$
to explain the observed spectrum. Indeed, molecular hydrogen 
requires higher pressure, at which the signature of abundant species 
like Na\,{\sc i}, K\,{\sc i}, H$_2$O, and possibly TiO 
should overcome the Rayleigh-scattering signature, except when they
have anomalously low abundances.

Because the cross section varies as $\propto a^6$ in the Rayleigh regime,
the scattering is largely dominated by the largest particles
in the size distribution.
Therefore, the typical sizes quoted above
must be considered as the size of the largest particles
in the distribution.

\section{Discussion}

This work assumes that the whole transit 
spectrum from 550 to 1050\,nm can be interpreted
by the absorption by a unique species. 
Alternatively, there could be a combination of absorptions 
by various species 
({\it e.g.}, Na\,{\sc i}, K\,{\sc i}, and H$_2$O)
in different parts of the spectrum 
to mimic a spectrum of Rayleigh scattering with the expected 
atmospheric temperature. This would be coincidental, 
but it cannot be excluded from the present data.
This question needs higher resolution data to be solved.

The temperature and pressure found above are mean
values integrated along the planetary limb, where
temperature is expected to vary from the pole to the equator
and from one side to the other.
A numerical integration of the absorption through the 
planetary atmosphere assuming a variation in temperature 
as a function of altitude of -60\,K per scale height (Burrows et al.\ 2007)
or as a function of longitude of 350\,K over 120 degrees (Knutson et al.\ 2007)
shows that temperatures found by solving Eq.~\ref{z_lambda}
are slightly overestimated by no more than about 3\% and 2\%, respectively.

If the measurements of planetary radius in the 550-1050\,nm
wavelength band are extrapolated to the
infrared, where Spitzer measurements are available, we can 
obtain a minimum effective radius due to Rayleigh scattering. 
From $R_p/R_*$=0.1564 at 750\,nm, we extrapolate that 
$R_p/R_*$ must be larger than 0.1539, 0.1531, and 0.1526
at 3.6\,$\mu$m, 5.8\,$\mu$m, and 8.0\,$\mu$m, respectively
(with $\pm$0.0003 (stat.) 0.0006 (syst.) error bars). 
These values are consistent with the Spitzer/IRAC measurements 
given by Knutson et al.\ (2007) and Ehrenreich et al.\ (2007), 
and with measurements by Beaulieu et al.\ (2007) 
at the limit of the upper error bars.

Rayleigh scattering has also been proposed to explain the
detection of polarized scattered light (Berdyugina et al.\ 2007).
However, this detection concerns the high upper atmosphere
and cannot be directly compared to the present work
on scattering deeper in the atmosphere.

Ground-based, high-resolution spectra allowed the detection 
of sodium in the atmosphere of HD\,18973b (Redfield et al.\ 2008).
Assuming a pressure at a given altitude, 
the measured absorption allows the determination
of the sodium abundance.
For that purpose, a fit of the sodium line shape 
is needed, and still has to be developed.
Alternatively, by assuming the sodium abundance, 
measuring sodium absorption should constrain
the pressure and allow for differentiating between the 
several carriers of the Rayleigh scattering proposed in
the present paper.
Presently, extinction high in the atmosphere by MgSiO$_3$ 
with its low condensation pressure of $\sim$10$^{-4}$\,bar appears a natural 
explanation for the absence of broad spectral signature
and is the preferred scenario.

\begin{acknowledgements}
We thank the anonymous referee for the constructive remarks.
We warmly thank J.-M.~D\'esert and G.~H\'ebrard for fruitful discussions.
\end{acknowledgements}

\end{document}